\documentclass[epj]{svjour}

\usepackage{graphicx}
\usepackage{fancyhdr}
\def\makeheadbox{{
 }}
\def\mytitle{My title} 
\def\myauthors{My name}  
\def\mytype{My type of session}
\def\mysession{My session}

\def\mytitle{Systematics and background suppression in the KATRIN experiment} 
\def\myauthors{K. Valerius}    
\def\mytype{Contributed Talk}    
\def\mysession{Cosmology and Astrophysics}

\begin{document}
\title{Systematics and background suppression in the KATRIN experiment}
\author{K. Valerius\inst{}\thanks{\emph{Email:} valerius@uni-muenster.de}%
~for the KATRIN collaboration
}                     
%
%
\institute{Institut f\"ur Kernphysik, Wilhelm-Klemm-Str. 9, Westf\"alische Wilhelms-Universit\"at M\"unster,\\ D-48149 M\"unster, Germany
}
%
\date{}
\abstract{Recent neutrino mass experiments at Mainz and Troitsk using tritium $\beta$-decay have reached their sensitivity potential, yielding upper limits of about $2\,\mathrm{eV}/c^2$ for $m(\bar{\nu}_e)$.
The KArlsruhe TRItium Neutrino ex\-peri\-ment (KATRIN), designed to reach a sensitivity of $m(\bar{\nu}_e) = 0.2\,\mathrm{eV}/c^2$ (90\% C.L.), will improve the signal rate by a factor of about $100$ with respect to previous experiments while maintaining the same low background level at an enhanced energy resolution of $0.93\,\mathrm{eV}$ of the spectrometer which is scaled up by a factor of 10 in linear dimensions. 
This low background rate can only be achieved by active and passive reduction of the background components induced by the spectrometer itself and in the detector region. Furthermore, sources of systematic errors such as energy losses inside the tritium source or fluctuations of the energy scale of the spectrometer need to be carefully controlled and analysed.
An overview of KATRIN's method to reduce the background rate and to determine the systematics as well
as the sensitivity on the neutrino mass will be presented.
} 
\maketitle
\section{Introduction} \label{intro}
Despite the results of neutrino oscillation experiments providing compelling evidence for non-zero neutrino masses, the absolute mass scale still remains undetermined. This question, which has a strong impact on both particle physics and cosmology, can be addressed in several, complementary ways, either via astro\-physi\-cal observations or by laboratory experiments. The direct neutrino mass determination relies on a precise measurement of the $\beta$ spectrum (tritium, $^{187}\textrm{Re}$) near its endpoint region. This kinematic approach, representing the only fully model-independent method of neutrino mass determination, will also be employed in the upcoming KATRIN experiment, which is being built on site of Forschungszentrum Karlsruhe, Germany. 

The spectrum $N$ of the kinetic electron energy $E_\mathrm{e}$ ({\it e.g.} for tritium $\beta$-decay, $^3\mathrm{H} \, \rightarrow \,^3\mathrm{He} + \bar{\nu}_\mathrm{e} + \mathrm{e}^-$) can be written as
\begin{eqnarray}
 N &=& K \cdot F(E_\mathrm{e}, Z+1) \cdot p_\mathrm{e} \cdot (E_\mathrm{e} + m_\mathrm{e}c^2)\cdot (E_0 - E_\mathrm{e}) \nonumber \\
   & & \cdot \sqrt{(E_0 - E_\mathrm{e})^2 - m^2(\nu_\mathrm{e})c^4},\label{equ:betaspec}
\end{eqnarray}
where $K$ contains the nuclear matrix element, $F$ is the Fermi function, $Z+1$ the charge number of the daughter nucleus, $p_e$ the electron momentum and $E_0$ the $\beta$ decay endpoint energy. Taking into account that the neutrino flavour eigenstates $\nu_\mathrm{e}$, $\nu_\mu$ and $\nu_\tau$ are related to the mass eigenstates $\nu_i$ via a mixing matrix $U$, and considering the smallness of the observed splitting of squared masses $\Delta m_{ij}^2$ from oscillation experiments, the neutrino mass observable in $\beta$-decay can be expressed as
\begin{equation}
 m^2(\nu_\mathrm{e}) = \sum_{i=1}^3 {|U_{ei}|^2 \cdot m^2(\nu_i)}.\label{equ:mixing}
\end{equation}
One can see from eq. (\ref{equ:betaspec}) that the influence of $m^2(\nu_\mathrm{e})$ on $N$ is largest in the region close to the $\beta$ endpoint energy where $E_0-E_\mathrm{e}$ is small. However, the decay rate in this energy range is only very low. Tritium as a $\beta$-decaying isotope is an ideal choice because of its low endpoint ($E_0 = 18.6~\mathrm{keV}$) and short half-life ($12.3~\mathrm{y}$). These advantages can best be exploited by using an electrostatic retardation spectrometer with the additional feature of magnetic adiabatic collimation\footnote{For details on the MAC-E filter technique (\underline{M}agnetic \underline{A}diabatic \underline{C}ollimation with an \underline{E}lectrostatic filter), see {\it e.g.} refs. \cite{picard} and \cite{KDR}.}, which allows to achieve a very good energy resolution and a high luminosity at the same time.

\section{Overview of the KATRIN experiment}
\begin{figure*}
\includegraphics[width=1.\textwidth,angle=0]{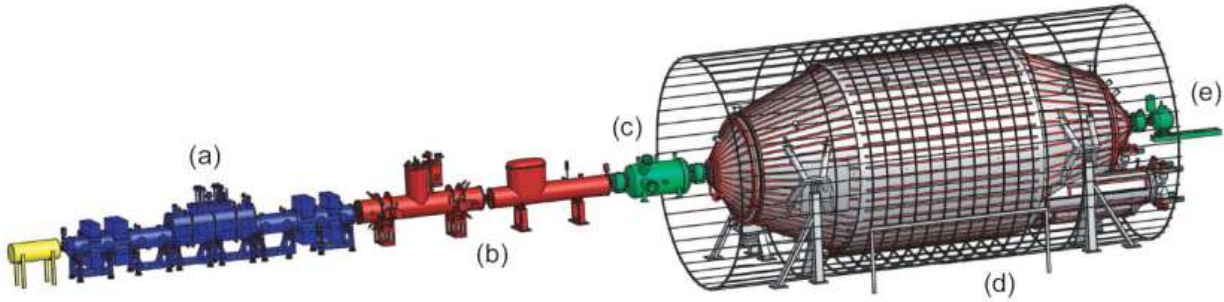}
\caption{Set-up of the KATRIN experiment: (a) windowless gaseous tritium source, (b) differential and cryo-pumping sections, (c) pre-spectrometer, (d) main spectrometer, (e) electron detector. The monitor spectrometer is not shown in this schematic view.}
\label{fig:katrinsetup}       
\end{figure*}

A schematic view of the KATRIN set-up is presented in fig. \ref{fig:katrinsetup}. One of the main components is the windowless gaseous tritium source (WGTS), consisting of a 10 m long tube of 90 mm diameter which contains $\mathrm{T}_2$ gas of high isotopic purity ($> 95\%$). The gas will be injected with a rate of $4.7\:\mathrm{Ci/s}$ at the center of the source and recovered at both ends to enter a closed loop. The column density ($\rho d = 5\cdot 10^{17}/\mathrm{cm}^2$), temperature ($T = 27\:\mathrm{K}$) and guiding magnetic field ($B = 3.6\:\mathrm{T}$) are among the operational parameters that need to be stabilised and controlled at a per mil level of precision. The electron transport system provides a magnetic field of $5.6\:\mathrm{T}$ to adiabadically guide the $\beta$-decay electrons towards the spectrometer, while at the same time reducing the amount of gas molecules by a differential pumping system using turbomolecular pumps. The subsequent cryo-pumping section is needed to further enhance the tritium retention factor and prevent $\mathrm{T}_2$ molecules from entering the spectrometer section. A pre-spectrometer serves as a means to reduce the energy interval of electrons reaching the main spectrometer to a region 200 to 300 eV below the endpoint $E_0$. Thus, only the part of the spectrum carrying the strongest imprint of a finite neutrino mass will be ana\-lysed with the large main energy filter which provides a relative resolution of $\Delta E/E = 1/20000$. Finally, after passing the main spectrometer, the electrons are detected by a segmented Silicon PIN diode with 148 pixels. More details on the experimental configuration can be found in ref. \cite{KDR}.
\section{Concepts for background reduction}
Since the count rate in the relevant signal region of the $\beta$ spectrum is so small, particular care needs to be taken in order to keep the background rate as low as possible. Considering the overall size of the experimental set-up, this clearly requires special measures. The linear dimensions of the main spectrometer, for instance, are about 10 times larger than the ones of the MAC-E filters previously operated at the neutrino mass experiments in Mainz and Troitsk. In this section, a short description of some aspects of the strategy for background reduction shall be given.
\begin{itemize}
 \item The whole spectrometer section will be kept at ultra-high vacuum ($p < 10^{-11}\:\mathrm{mbar}$) in order to minimise the scattering of electrons on residual gas and thus the possibility of io\-ni\-sa\-tion-induced background. Both spectrometers are equipped with a heating system allowing bake-out at temperatures up to $350^\circ\,\mathrm{C}$.
 \item The retardation potential of $\sim 18.6\:\mathrm{kV}$ will be applied directly to the stainless steel vacuum vessels of the spectrometers. Secondary electrons created by radioactive contaminations in the material or by impinging cosmic muons have a small, but finite chance of reaching the inner volume of the spectrometers, from where they can follow magnetic field lines through the spectrometer and onto the detector. Electrons starting with low energies will get accelerated by the negative potential on the vessel electrode to an energy practically indistinguishable from signal electrons within the energy resolution of the detector. A quasi-massless wire electrode placed on a more negative potential relative to the tank will be implemented in order to prevent those secondary electrons from reaching the inner volume and in particular the detector. Experience at the Mainz MAC-E filter \cite{flatt} has demonstrated that a potential difference of about $100\:\mathrm{V}$ is sufficient to reduce the background level by a factor of 10 into the mHz region. A double-layer inner electrode divided into $\sim 240$ modules will be installed inside the main spectrometer, covering its full inner surface area of $690\mathrm{m}^2$. The purpose of using two wire layers with $100\:\mathrm{V}$ potential difference to the next layer (or to the vessel wall, respectively) is to further enhance the shielding factor. In addition, the two-layer design allows to hide most parts of the solid-metal mounting structures, which could in principle form a source of secondary electrons themselves, behind the innermost layer.
 \item Arising from its combination of electric and magnetic fields, the MAC-E filter naturally offers conditions for the trapping of charged particles. These traps can be a source of additional background, since trapped particles have a longer pathlength and, depending on their energy, can cause ionisation (even in an ultra-high vacuum environment). Therefore, the avoidance of such trapping conditions has to be among the design criteria for the vacuum tank and inner electrode system. In addition, the inner electrode of the main spectrometer will consist of two separate halves which can be used to apply an electric dipole field that helps to remove stored particles by means of the $\vec{E}\times\vec{B}$ drift motion.
 \item The electron detector, finally, will be situated inside a low-activity passive shielding and an active veto system. The option of post-accelerating the electrons by a potential of $-25\:\mathrm{kV}$ is considered in order to shift the signal to an energy region with lower intrinsic background.

\end{itemize}

\section{Sources of systematic uncertainties}
\begin{figure}
\centering
\includegraphics[width=0.45\textwidth,angle=0]{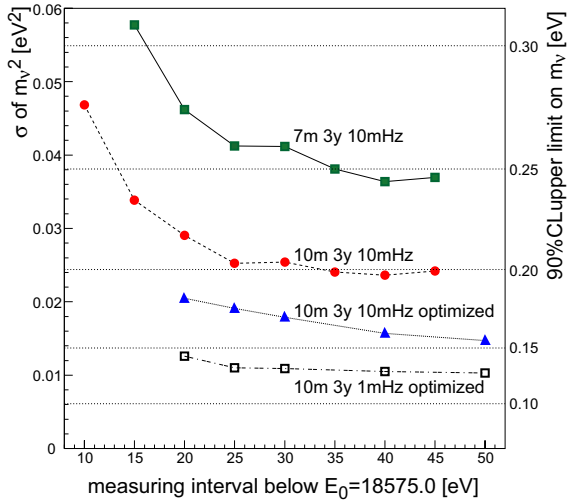}
\caption{Statistical uncertainty on the observable $m_\nu^2$ and 90\% C.L. upper limit on $m_\nu$ depending on different analyisis intervals. The top curve shows the values stated in the Letter of Intent \cite{LoI}. Since then, significant improvement has been achieved by enlarging the diameter of the tritium source from $75\:\mathrm{mm}$  to $90\:\mathrm{mm}$ (requiring the diameter of the main spectrometer to be raised from $7\:\mathrm{m}$ to $10\:\mathrm{m}$). Reducing the assumed background rate of $10\:\mathrm{mHz}$ by a factor of 10 could improve the sensitivity even further.}
\label{fig:statistics}       
\end{figure}

KATRIN's sensitivity aim is based on detailed studies identifying and evaluating the influence of a number of systematic uncertainties, as presented in an extensive discussion in ref. \cite{KDR}. A summary of the main sources of systematic errors and the measures foreseen by the KATRIN collaboration to keep their influence as low as possible is given in the list below.
\begin{enumerate}
 \item \emph{Inelastic scattering of electrons inside the WGTS.} The energy loss function needs to be determined via deconvolution techniques from dedicated measurements of the response function (using an electron gun as a source of monoenergetic electrons). 
 \item \emph{Fluctuations of the column density in the WGTS.} The temperature of the tritium source needs to be stabilised on a $0.1\%$ level. Two methods to check for fluctuations in the column density consist of (a) repeated electron gun measurements to determine the ratio of scattered to non-scattered electrons, and (b) on-line mass spectrometry to determine the isotopic composition combined with a measurement of the total $\beta$-activity of the WGTS.
  \item \emph{HV stability of the retarding potential.} Unrecognised shifts of the retarding potential constitute an important source of systematic uncertainty, as demonstrated in a study presented in ref. \cite{kaspar}. The retarding potential will be monitored by a custom-built precision high-voltage divider read by a precision digital voltmeter. A third MAC-E filter (the former spectrometer of the Mainz neutrino mass experiment), connected to the same high voltage as the main spectrometer, will form a secondary beam line. This 'monitor spectrometer' allows continuous control of the absolute energy scale, making use of dedicated calibration sources based on monoenergetic atomic and nuclear standards (such as conversion electrons from $^{83\mathrm{m}}\mathrm{Kr}$ or Co photoelectrons generated from $^{241}\mathrm{Am}$ $\gamma$-rays). These sources need to be prepared in a way that assures long-term stability (and reproducibility) on a ppm level.
\item \emph{Corrections to the transmission function.} The transmission function of the MAC-E filter depends on the relative values of the electric potential $U$ and the magnetic field $B$. Unavoidable inhomogeneities of $U$ and $B$ across the diameter of the spectrometer will be mapped onto the sensitive area of the electron detector in form of a varying shape of the transmission function on each pixel. These small corrections need to be determined in dedicated measurements using an electron gun or a $^{83\mathrm{m}}\mathrm{Kr}$ test source at the main spectrometer before the start of tritium runs. 
 \item \emph{Potential variations in the WGTS due to a build-up of space charges.} With a $\beta$-activity inside the WGTS as high as $10^{11}\:\mathrm{Bq}$, ionising collisions between the high-energy electrons and gas molecules are likely to play a significant role. On average, 15 secondary electron-ion pairs are expected to be created by each $\beta$-electron. As a consequence, the gas inside the WGTS will contain a considerable amount of $\mathrm{T}^+$, $\mathrm{He}^+$, $(^3\mathrm{HeT})^+$, $\mathrm{T}_2^+$ and other positive ions, forming a quasi-neutral plasma with the secondary electrons from the ionisation. The energy distribution of these secondary electrons which determines the magnitude of the space charge potential variation has been the subject of Monte Carlo simulations \cite{nasto}. The outcome of these studies implies that uncompensated space charge variations inside the source should be of the order of $10\:\mathrm{mV}$. If necessary, the neutrality of the plasma could be enhanced by adding low-energy electrons from the rear side of the WGTS.
 \item \emph{Electronic final state distribution of daughter mo\-le\-cules.} The theoretical description of the $\beta$ spectrum as given in eq. (\ref{equ:betaspec}) needs to be altered by taking into account excitations of the daughter molecules. As the electronic structure of the molecules in question is rather simple, reliable quantum-chemical calculations are available. In view of KATRIN's sensitivity aim and the increased requirements with respect to previous ana\-ly\-ses of the tritium $\beta$ spectrum, a new investigation of the molecular effects has been carried out \cite{doss}, with emphasis on the final state distribution of the six lowest electronic states in $^3\mathrm{HeT}^+$, $^3\mathrm{HeD}^+$ and $^3\mathrm{HeH}^+$. This limitation is justified by the small energy range -- about $30 \:\mathrm{eV}$ below the endpoint $E_0$ -- envisioned as an evaluation interval for the neutrino mass determination in KATRIN. As the first electronic excited state in $^3\mathrm{HeT}^+$ has an excitation energy of $27\;\mathrm{eV}$, rotational-vibrational excitations of the electronic ground state with their mean excitation energy of $1.7\:\mathrm{eV}$ play a more important role.
\end{enumerate}
Each of the considered sources of systematic uncertainties is estimated to give a contribution of about $\sigma_\mathrm{syst}(m_\nu^2) \leq 0.007\;\mathrm{eV}^2/c^4$, thus adding up quadratically to a total systematic error of $\sigma_\mathrm{syst,tot}(m_\nu^2) \leq 0.017\;\mathrm{eV}^2/c^4$. This number is to be compared to the expected statistical error. After introducing several improvements, such as an increased luminosity due to a larger diameter of source and main spectrometer, enhanced isotopic purity of the $\mathrm{T}_2$ gas, and an optimised distribution of measurement time at various settings of the retarding potential (compare fig. \ref{fig:statistics}), the statistical uncertainty after 3 years of full data taking is estimated to be $\sigma_\mathrm{stat}(m_\nu^2) = 0.018\;\mathrm{eV}^2/c^4$. Here, an analysing interval of $[E_0 - 30\:\mathrm{eV},\;E_0 + 5\:\mathrm{eV}]$ was assumed. This means that after 3 effective beam years, statistical and systematic errors will contribute about equally, yielding a total uncertainty of 
\begin{equation}
\sigma_\mathrm{tot}(m_\nu^2) = 0.025\:\mathrm{eV}^2/c^4. 
\end{equation}
 Assuming $m_\nu = 0$, this uncertainty corresponds to an upper limit of 
\begin{equation}
L(90\%\:\mathrm{C.L.}) = \sqrt{1.64\cdot \sigma_\mathrm{tot}} = 0.2\:\mathrm{eV}/c^2.
\end{equation} 
A neutrino mass of $m_\nu = 0.30 \;\mathrm{eV}/c^2$ $(0.35 \;\mathrm{eV}/c^2)$ could be detected with a $3\sigma$ ($5\sigma$) significance.

In a recent analysis \cite{bayes-freq} it was shown that this estimate using the frequentist approach is a rather conservative one, and by applying a Bayesian inference method instead, the authors arrived at an estimated sensitivity of $0.17\:\mathrm{eV}/c^2~(90\%\:\mathrm{C.L.})$.

\section{Status and schedule of the KATRIN experiment}
Installation works to set up the KATRIN experiment are presently under way at Forschungszentrum Karls\-ruhe, Germany. The new building facilities next to the Tritium Laboratory Karlsruhe (TLK) are already available, one of them housing the main spectrometer vessel, which has been delivered to Karlsruhe in November 2006. Vacuum tests with the main spectrometer tank (including a bake-out cycle up to a temperature of $350^\circ\,\mathrm{C}$) yielded a preliminary outgassing rate of $1\times 10^{-12}\;\mathrm{mbar\, l\,s^{-1}\, cm^{-2}}$, thus successfully meeting the challenging KATRIN requirements. The inner electrode system for the main spectrometer is now being assembled at M\"unster University and will be shipped to Karlsruhe and installed in early 2008. The second major component, the WGTS, is presently being manufactured at an external company and delivery is expected in 2009.

Commissioning of the full set-up and the start of data taking is scheduled for 2010. A total measuring time of 5 years will be necessary to collect the 3 years of effective beam time required to reach KATRIN's design sensitivity.

\begin{section}*{Acknowledgements}
The author's work for the KATRIN collaboration is supported by the German Federal Ministry of Education and Research (BMBF).
\end{section}
%

%

\begin{thebibliography}{999}
%
%
\bibitem{picard}
A. Picard et al., Nucl. Instr. Meth. B {\bf 63}, (1992) 345.
\bibitem{KDR}
J. Angrik et al., FZKA Scientific Report 7090 (2005), {\it http://bibliothek.fzk.de/zb/berichte/FZKA7090.pdf}.
\bibitem{flatt}
B. Flatt, PhD thesis, Universty of Mainz (2004).
\bibitem{kaspar}
J. Kaspar et al., Nucl. Instr. Meth. A {\bf 527}, (2005) 423.
\bibitem{nasto}
A. F. Nastoyashchii et al., Fusion Science and Technology 48, (2005) 743.
\bibitem{doss}
N. Doss, J. Tennyson, A. Saenz and S. Jonsell, Phys. Rev. C {\bf 73}, (2006) 025502.
\bibitem{LoI}
A. Osipowicz et al., FZKA Scientific Report 6691 (2001), {\tt hep-ex/0109033}.
\bibitem{bayes-freq}
O. Host, O. Lahav, F. B. Abdalla and K. Eitel, arXiv:0709.1317v1 [hep-ph] subm. to Phys. Rev. D.

\end{thebibliography}
\end{document}